\documentclass[proceedings]{stacs}
\stacsheading{2010}{395-404}{Nancy, France}
\firstpageno{395}

\usepackage{amsfonts}
\usepackage{amsmath}
\usepackage{amssymb}
\usepackage{enumerate}

\usepackage{amssymb,latexsym,amsmath,ifthen}

\newcommand{\sinf}{\Sigma^{\infty}}
\newcommand{\sfin}{\Sigma^{*}}

\newcommand{\PROOF}{\begin{proof}}

\newcommand{\QED}{\end{proof}}

\newcommand{\bval}[1]{[\![ #1 ]\!]}

\newcommand{\binary}{\{0,1\}^*}

\newcommand{\N}{\mathbb{N}}
\newcommand{\Z}{\mathbb{Z}}

\newcommand{\p}{{\mathrm{p}}}
\newcommand{\ptwo}{{\p_{\thinspace\negthinspace_2}}}

\newcommand{\mup}{\mu_{\mathrm{p}}}

\newcommand{\co}[1]{\mathrm{co}#1}

\newcommand{\bininf}{\{0,1\}^{\infty}}

\newcommand{\TIME}{\mathrm{TIME}}
\newcommand{\DTIME}{\mathrm{DTIME}}

\newcommand{\NP}{{\ensuremath{\mathrm{NP}}}}
\newcommand{\UP}{\mathrm{UP}}
\newcommand{\coNP}{\co{\NP}}

\newcommand{\PSPACE}{{\rm PSPACE}}
\renewcommand{\P}{\ensuremath{{\mathrm P}}}

\newcommand{\disjNP}{\mathrm{disjNP}}
\newcommand{\disjEXP}{\mathrm{disjEXP}}

\newcommand{\EXP}{{\rm EXP}}

\newcommand{\calC}{{\mathcal C}}

\newcommand{\supp}{\mathrm{supp}}
\newcommand{\ind}{\mathrm{ind}}

\newcommand{\nextd}{\mathrm{next}}

\renewcommand{\include}{\input}

\newcommand{\alf}{\{0, 1\}}

\newcommand{\bin}[1]{\left\{0, 1\right\}^{#1}}

\newcommand{\mcc}{{\mathcal{C}}}

\newcommand{\set}[2]{\left\{ {#1} \: \Big| \: {#2} \right\} }

\newcommand{\sinfd}{S^{\infty}[d]}

\newcommand{\TIMEdnk}{{\mathrm{TIME}(2^{n^k})}}

\newtheorem{construction}[thm]{Construction}

\newenvironment{example*}[1]
{ {\noindent {\bf Example #1.}} } {  }
\newenvironment{claim*}[1]
{ {\noindent {\bf Claim #1.}} } {  }
\newenvironment{theorem*}
{ {\noindent {\bf Theorem.}} } {  }

\numberwithin{equation}{section} \numberwithin{thm}{section}

\newenvironment{notation}
{ {\noindent {\bf Notation.}} } {
 }

\begin{document}

\title[Inseparability and Strong Hypotheses for Disjoint NP Pairs]{Inseparability and Strong Hypotheses for Disjoint NP Pairs}

\author[lab1]{L. Fortnow}{Lance Fortnow}
\address[lab1]{Northwestern University, EECS Department, Evanston, Illinois, USA.
}  
\email{fortnow@eecs.northwestern.edu}  

\author[lab2]{J. H. Lutz}{Jack H. Lutz}
\address[lab2]{Department of Computer Science, Iowa State University, Ames, IA 50011 USA.  }
\email{lutz@cs.iastate.edu}  

\author[lab3]{E. Mayordomo}{Elvira Mayordomo}
\address[lab2]{Departamento de Inform\'atica e Ingenier\'ia de Sistemas, Instituto de Investigaci\'on en
Ingenier\'ia de Arag\'on, Mar\'ia de Luna 1, Universidad de Zaragoza, 50018 Zaragoza, SPAIN.}
\email{elvira\ at\ unizar.es}  

\thanks{Thanks: Fortnow's research supported in part by NSF grants
CCF-0829754 and DMS-0652521. Lutz's research supported in part by National Science Foundation Grants 0344187, 0652569, and 0728806. Mayordomo's
research supported in part by Spanish Government MICINN
   Project
   TIN2008-06582-C03-02.} 



\begin{abstract}
  \noindent This paper investigates the existence of inseparable disjoint
pairs of NP languages and related strong hypotheses in computational complexity.  Our main theorem says that, if NP does not have measure 0 in EXP,
then there exist disjoint pairs of NP languages that are P-inseparable, in fact TIME(2(n k))-inseparable. We also relate these conditions to strong
hypotheses concerning randomness and genericity of disjoint pairs.
\end{abstract}

\maketitle

\newcounter{fig1}\newcounter{fig2}

\setcounter{fig1}{2} \setcounter{fig2}{1}
\section{Introduction}

The main objective of complexity theory is to assess the intrinsic
difficulties of naturally arising computational problems.  It is
often the case that a problem of interest can be formulated as a
decision problem, or else associated with a decision problem of
the same complexity, so much of complexity theory is focused on
decision problems.  Nevertheless, other types of problems also
require investigation.

This paper concerns {\sl promise problems\/}, a natural
generalization of decision problems introduced by Even, Selman,
and Yacobi \cite{EvSeYa84}. A decision problem can be formulated
as a set $A \subseteq \binary$, where a solution of this problem
is an algorithm, circuit, or other device that {\sl decides\/}
$A$, i.e., tells whether or not an arbitrary input $x \in \binary$
is an element of $A$.  In contrast, a promise problem is
formulated as an ordered pair $(A,B)$ of disjoint sets $A, B
\subseteq \binary$, where a solution is an algorithm or other
device that decides {\sl any\/} set $S \subseteq \binary$ such
that $A \subseteq S$ and $B \cap S = \emptyset$.  Such a set $S$
is called a {\sl separator\/} of the disjoint pair $(A,B)$.
Intuitively, if we are promised that every input will be an
element of $A \cup B$, then a separator of $(A,B)$ enables us to
distinguish inputs in $A$ from inputs in $B$.  Since each decision
problem $A$ is clearly equivalent to the promise problem $(A,
A^c)$, where $A^c = \binary -A$ is the complement of $A$, promise
problems are, indeed, a generalization of decision problems.

A {\sl disjoint NP pair\/} is a promise problem $(A,B)$ in which
$A, B \in \NP$.  Disjoint NP pairs were first investigated by
Selman and others in connection with public key cryptosystems
\cite{EvSeYa84,GroSel88,Selm89,HomSel92}.  They were later
investigated by Razborov \cite{Razb94} as a setting in which to
prove the independence of complexity-theoretic conjectures from
theories of bounded arithmetic. In this same paper, Razborov
established a fundamental connection between disjoint NP pairs and
propositional proof systems. Propositional proof systems had been
used by Cook and Reckhow \cite{CooRec79} to characterize the NP
versus co-NP problem. Razborov \cite{Razb94} showed that each
propositional proof system has associated with it a canonical
disjoint NP pair and that important questions about propositional
proof systems are thereby closely related to natural questions
about disjoint NP pairs.  This connection with propositional proof
systems has motivated more recent work on disjoint NP pairs by
Gla\ss{}er, Selman, Sengupta, and Zhang \cite{GlSeSeZh04,
GlaSelSen05,GlaSelZha07a,GlaSelZha07b}.  It is now known that the
degree structure of propositional proof systems under the natural
notion of proof simulation is identical to the degree structure of
disjoint NP pairs under reducibility of separators
\cite{GlaSelZha07a}. Much of this recent work is surveyed in
\cite{GlaSelZha06}.  Goldreich \cite{Gold06b} gives a recent
survey of promise problems in general.

Our specific interest in this paper is the existence of disjoint NP pairs that are P-inseparable, or even $\TIMEdnk$-inseparable. As the terminology
suggests, if $\calC$ is a class of decision problems, then a disjoint pair is {\sl $\calC$-inseparable\/} if it has no separator in $\calC$.  The
existence of P-inseparable disjoint NP pairs is a strong hypothesis in the sense that (1) it clearly implies $\P \ne \NP$, and (2) the converse
implication is not known (and fails relative to some oracles \cite{HomSel92}). It is clear that $\P \ne \NP\cap \coNP$ implies the existence of
P-inseparable disjoint NP pairs, and Grollmann and Selman \cite{GroSel88} proved that $\P \ne \UP$ also implies the existence of P-inseparable
disjoint NP pairs.

The hypothesis that NP is a non-measure 0 subset of EXP, written
$\mu(\NP\mid \EXP)\ne 0$, is a strong hypothesis in the above
sense. This hypothesis has been shown to have many consequences
not known to follow from more traditional hypotheses such as $\P
\ne \NP$ or the separation of the polynomial-time hierarchy into
infinitely many levels.  Each of these known consequences has
resolved some pre-existing complexity-theoretic question in the
way that agreed with the conjecture of most experts.  This
explanatory power of the $\mu(\NP\mid \EXP)\ne 0$ hypothesis is
discussed in the early survey papers
\cite{QSET,AmbMay97,LutMayTPRBM} and is further substantiated by
more recent papers listed at \cite{Hitch-bib} (and too numerous to
discuss here).  In several instances, the discovery that
$\mu(\NP\mid \EXP)\ne 0$ implies some plausible conclusion has led
to subsequent work deriving the same conclusion from some weaker
hypothesis, thereby further illuminating the relationships among
strong hypotheses.

Our main theorem states that, if NP does not have measure zero in
EXP, then, for every positive integer $k$, there exist disjoint NP
pairs that are $\TIMEdnk$-inseparable.  Such pairs are {\sl a
fortiori\/} P-inseparable, but the conclusion of our main theorem
actually gives {\sl exponential\/} lower bounds on the
inseparability of some disjoint NP pairs.  These are the lower
bounds that most experts conjecture to be true, even though an
unconditional proof of such bounds may be long in coming.

The proof of our main theorem combines known closure properties of
NP with the randomness that the $\mu(\NP\mid \EXP)\ne 0$
hypothesis implies must be present in NP to give an explicit
construction of a disjoint NP pair that is $\TIMEdnk$-inseparable.
(Technically, this is an overstatement. The last step of the
``construction'' is the removal of a finite set whose existence we
prove, but which we do not construct.)  The details are perhaps
involved, but we preface the proof with an intuitive motivation
for the approach.

We also investigate the relationships between the two
strong hypotheses in our main theorem (i.e., its hypothesis
and its conclusion) and strong hypotheses involving the
existence of disjoint NP pairs with randomness and
genericity properties. Roughly speaking (i.e., omitting
quantitative parameters), we show that the existence of
disjoint NP pairs that are random implies both the
$\mu(\NP\mid \EXP)\ne 0$ hypothesis and the existence of
disjoint NP pairs that are generic in the sense of
Ambos-Spies, Fleischhack, and Huwig \cite{AmFlHu87}.  We
also show that the existence of such generic pairs implies
the existence of disjoint NP pairs that are
$\TIMEdnk$-inseparable. Taken together, these results give
the four implications at the top of Figure \arabic{fig2}.
(The four implications at the bottom are well known.)  We
prove that three of these implications cannot be reversed
by relativizable techniques, and we conjecture that this
also holds for the remaining implication.

\section{Preliminaries}

We write $\N$ for the set of nonnegative integers and $\Z^{+}$ for
the set of (strictly) positive integers. The {\sl Boolean value\/}
of an assertion $\phi$ is $\bval{\phi}=$ {\sl if $\phi$ then 1
else 0}. All logarithms here are base-2.

We write $\lambda$ for the empty string, $|w|$ for the length of a
string $w$, and $s_0, s_1, s_2, \ldots$ for the standard
enumeration of $\binary$. The index of a string $x$ is the value
$\ind(x)\in\N$ such that $s_{\ind(x)}=x$. We write $\nextd(x)$ for
the string following $x$ in the standard enumeration, i.e.,
$\nextd(s_n)=s_{n+1}$. More generally, for $k\in\N$, we write
$\nextd^k$ for the $k$-fold composition of next with itself, so
that $\nextd^k(s_n)=s_{n+k}$.

A {\sl Boolean function\/} is a function $f: \bin{m}\to \alf$ for
some $m\in\N$. The {\sl support\/} of such a function $f$ is
$\supp(f)=\set{x\in\bin{m}}{f(x)=1}$.

We write $w[i]$ for the $i^{\mathrm{th}}$ symbol in a string $w$
and $w[i..j]$ for the string consisting of the $i^{\mathrm{th}}$
through $j^{\mathrm{th}}$ symbols. The leftmost symbol of $w$ is
$w[0]$, so that $w=w[0..|w|-1]$. For (infinite) sequences
$S\in\sinf$, the notations $S[i]$ and $S[i..j]$ are defined
similarly. A string $w\in\sfin$ is a prefix of a string or
sequence $x\in\sfin\cup \sinf$, and we write $w\sqsubseteq x$, if
there is a string or sequence $y\in\sfin\cup \sinf$ such that
$wy=x$. A {\sl language}, or {\sl decision problem,} is a set
$A\subseteq \binary$. We identify each language $A$ with the
sequence $A\in\bininf$ defined by $A[n]=\bval{s_n\in A}$ for all
$n\in\N$. If $A$ is a language, then expressions like $\lim_{w\to
A}f(w)$ refer to prefixes $w\sqsubseteq A$, e.g., $\lim_{w\to A}
f(w)= \lim_{n\to \infty}f(A[0..n-1])$.

A {\sl martingale\/} is a function $d: \{0,1\}^*\to [0, \infty)$
satisfying \begin{equation}\label{equ21}
d(w)=\frac{d(w0)+d(w1)}{2}\end{equation} for all $w\in\binary$.
Intuitively, $d$ is a {\sl strategy for betting\/} on the
successive bits of a sequence $S\in\bininf$: The quantity $d(w)$
is the amount of money that the gambler using this strategy has
after $|w|$ bets if $w\sqsubseteq S$. Condition (\ref{equ21}) says
that the payoffs are fair.

A martingale $d$ {\sl succeeds\/} on a language $A\subseteq\binary$, and we write $A\in\sinfd$, if \newline $\limsup_{w\to A}d(w)=\infty$. If
$t:\N\to\N$, then a martingale $d$ is ({\sl exactly}) $t(n)$-{\sl computable\/} if its values are rational and there is an algorithm that computes
each $d(w)$ in $t(|w|)$ time. A martingale is p-{\sl computable\/} if it is $n^k$-computable for some $k\in\N$, and it is $\ptwo$-computable if it is
$2^{(\log n)^k}$-computable for some $k\in\N$.

\begin{definition}\cite{AEHNC} Let $X$ be a set of languages, and
let $R$ be a language.
\begin{enumerate}
\item $X$ has p{\sl -measure 0}, and we write $\mup(X)=0$, if
there is a p-computable martingale $d$ such that $X\subseteq
\sinfd$. The condition $\mu_{\ptwo}(X)=0$ is defined analogously.
\item $X$ has {\sl measure 0 in\/} EXP, and we write $\mu(X\mid
\EXP)=0$, if $\mu_{\ptwo}(X\cap \EXP)=0$. \item $R$ is p-{\sl
random\/} if $\mup(\{R\})\ne 0$, i.e., if there is no p-computable
martingale that succeeds on $R$. Similarly, $R$ is $t(n)$-{\sl
random\/} if no $t(n)$- computable martingale succeeds on $R$.

\end{enumerate}\end{definition}

It is well known that these definitions impose a nontrivial
measure structure on EXP \cite{AEHNC}. For example, $\mu(\EXP\mid
\EXP)\ne 0$.

We use the following fact in our arguments.

\begin{lemma}\cite{rAmTeZh97,JUEDES1995a}\label{lem21} The following five
conditions are equivalent.
\begin{enumerate}
\item $\mu(\NP\mid \EXP)\ne 0$. \item $\mup(\NP)\ne 0$. \item
$\mu_{\ptwo}(\NP)\ne 0$. \item There exists a p-random
language $R\in\NP$. \item For every $k\ge 2$, there exists
an $2^{\log n^k}$-random language $R\in\NP$.
\end{enumerate}

\end{lemma}

Finally, we note that $\mu(\P\mid \EXP)=0$ \cite{AEHNC}, so
$\mu(\NP\mid \EXP)\ne 0$ implies $\P\ne\NP$.

\section{Inseparable Disjoint NP Pairs and the Measure of NP}\label{sec4}

This section presents our main theorem, which says that, if NP does not have measure 0 in EXP, then there are disjoint NP pairs that are
P-inseparable. In fact, for each $k\in\N$, there is a disjoint NP pair that is $\TIMEdnk$-inseparable.

It is convenient for our arguments to use a slight variant of the separability notion.

\begin{definition} Let $(A,B)$ be a pair of (not necessarily disjoint) languages, and let $\mcc$ be a class of languages.
\begin{enumerate}
\item A language $S\subseteq\binary$ {\sl almost separates\/} $(A,B)$ if there is a finite set $D\subseteq\binary$ such that $S$ separates $(A-D,
B-D)$. \item We say that $(A, B)$ is {\sl $\mcc$-almost separable \/} if there is a language $S\in\mcc$ that almost separates $(A, B)$.
\end{enumerate}
\end{definition}

\begin{observation}\label{obs41} If a pair $(A, B)$ is not $\mcc$-almost separable, then $(A-D, B-D)$ is $\mcc$-inseparable for every finite set $D$.
\end{observation}

Before proving our main theorem, we sketch the intuitive idea of the proof. We want to construct a disjoint NP pair $(A, B)$ that is P-inseparable.
Our hypothesis, that NP does not have measure 0 in EXP, implies that NP contains a language $R$ that is p-random. Since we are being intuitive, we
ignore the subtleties of p-randomness and regard $R$ as a sequence of independent, fair coin tosses (with the $n^{\mathrm{th}}$ toss heads iff
$s_n\in R$) that just happens to be in NP. If we use these coins to randomly put strings in $A$ or $B$ but not both, we can count on the randomness
to thwart any would-be separator in P.

The challenge here is that, if we are to deduce $A, B \in \NP$ from $R\in\NP$, we must make the conditions ``$s_n\in A$'' and ``$s_n\in B$'' depend
on the coin tosses in a {\sl monotone\/} way; i.e., adding a string to $R$ must not move a string out of $A$ or out of $B$.

This monotonicity restriction might at first seem to
prevent us from ensuring that $A$ and $B$ are disjoint.
However, this is not the case. Suppose that we decide
membership of the $n^{\mathrm{th}}$ string $s_n$ in $A$ and
$B$ in the following manner. We toss $2\log n$ independent
coins. If the first $\log n$ tosses all come up heads, we
put $s_n$ in $A$. If the second $\log n$ tosses all come up
heads, we put $s_n$ in $B$. If our coin tosses are taken
from $R$, which is in NP, then $A$ and $B$ will be in NP.
Each string $s_n$ will be in $A$ with probability
$\frac{1}{n}$, in $B$ with probability $\frac{1}{n}$, and
in $A\cap B$ with probability $\frac{1}{n^2}$. Since
$\sum_{n=1}^{\infty}\frac{1}{n}$ diverges and
$\sum_{n=1}^{\infty}\frac{1}{n^2}$ converges, the first and
second Borel-Cantelli lemmas tell us that $A$ and $B$ are
infinite and $A\cap B$ is finite. Since $A\cap B$ is
finite, we can subtract it from $A$ and $B$, leaving two
disjoint NP languages that are, by the randomness of the
construction, P-inseparable.

What prevents this intuitive argument from being a proof sketch is
the fact that the language $R$ is not truly random, but only
p-random. The proof that $A\cap B$ is finite thus becomes
problematic. There is a resource-bounded extension of the first
Borel-Cantelli lemma \cite{AEHNC} that works for p-random
sequences, but this extension requires the relevant sum of
probabilities to be p-convergent, i.e., to converge much more
quickly than $\sum_{n=1}^{\infty}\frac{1}{n^2}$.

Fortunately, in this particular instance, we can achieve our objective without p-conver\-gen\-ce or the (classical or resource-bounded)
Borel-Cantelli lemmas. We do this by modifying the above construction. Instead of putting the $n^{\mathrm{th}}$ string into each language with
probability $\frac{1}{n}$, we put each string $x$ into each of $A$ and $B$ with probability $2^{-|x|}$ so that $x$ is in $A\cap B$ with probability
$2^{-2|x|}$. By the Cauchy condensation test, the relevant series have the same convergence behavior as those in our intuitive argument, but we can
now replace slow approximations of tails of $\sum_{n=1}^{\infty}\frac{1}{n^2}$  with fast and exact computations of geometric series.

We now turn to the details.

\begin{construction}\label{con42}
\begin{enumerate}
\item Define the functions $u, v: \binary \to\binary$ by the recursion
\[\begin{array}{l}u(\lambda)=\lambda,\\
v(x)=\nextd^{|x|}(u(x)),\\
u(\nextd(x)=\nextd^{|x|}(v(x)).\end{array}\] \item For each
$x\in\binary$, define the intervals \[I_x=[u(x), v(x)), \
J_x=[v(x), u(next(x))).\] \item For each $R\subseteq\binary$,
define the languages
\[\begin{array}{l}A^{+}(R)=\set{x}{I_x\subseteq R},\ B^{+}(R)=\set{x}{J_x\subseteq R},\\
A(R)= A^{+}(R) - B^{+}(R),\ B(R)=B^{+}(R)- A^{+}(R).\end{array}\]
\end{enumerate}
\end{construction}

Note that each $|I_x|=|J_x|=|x|$. Also,
$I_{\lambda}=J_{\lambda}=\emptyset$ (so $\lambda \in A^{+}(R)\cap
B^{+}(R)$), and
\[I_0<J_0<I_1<J_1<I_{00}<J_{00}<I_{01}<\ldots,\]
with these intervals covering all of $\binary$.

A routine witness argument gives the following.

\begin{observation}\label{obs43}\begin{enumerate}\item If $R\in\NP$, then $A^{+}(R), B^{+}(R)\in \NP$.
\item If $R\in\NP$ and $|A^{+}(R)\cap B^{+}(R)|<\infty$, then $(A(R), B(R))$ is a disjoint NP pair.\end{enumerate}\end{observation}

We now prove two lemmas about Construction \ref{con42}.

\begin{lemma}\label{lem44} Let $k\in\N$. If $R\subseteq \binary$ is $2^{(\log n)^{k+2}}$- random, then $(A^{+}(R), B^{+}(R))$ is not
$\TIMEdnk$-almost separable.\end{lemma}

\begin{lemma}\label{lem45} If $R\subseteq\binary$ is p-random, then $|A^{+}(R)\cap B^{+}(R)|<\infty$.\end{lemma}

We now have what we need to prove our main result.

\begin{theorem}{(main theorem)}\label{the46}  If NP does not have measure 0 in EXP, then, for every $k\in\Z^{+}$, there is a disjoint NP pair that is
$\TIMEdnk$-inseparable, hence certainly P-inseparable. \end{theorem}

\begin{proof} Assume that $\mu(\NP\mid\EXP)\ne 0$, and let $k\in\N$. Then, by Lemma \ref{lem21}, there is a  $2^{(\log n)^{k+2}}$-random language
$R\in\NP$. By Lemma \ref{lem44}, the pair $(A^{+}(R), B^{+}(R))$ is not $\TIMEdnk$-almost separable. Since $R$ is certainly p-random, Lemma
\ref{lem45} tells us that $|A^{+}(R)\cap B^{+}(R)|<\infty$. It follows by Observation \ref{obs43}\ that $(A(R), B(R))$ is a disjoint NP pair, and it
follows by Observation \ref{obs41}\ that $(A(R), B(R))$ is $\TIMEdnk$-inseparable. \end{proof}

\section{Genericity and Measure of Disjoint NP Pairs}\label{sec5}

In this section we introduce the natural notions of resource-bounded measure and genericity for disjoint pairs and relate them to the existence of
P-inseparable pairs in NP. We compare the different strength hypothesis on the
 measure and genericity of NP and disjNP establishing all the relations in Figure \arabic{fig2}.

\begin{notation}  Each disjoint pair $(A,B)$ will be coded as an infinite sequence $T\in \{-1,0,1\}^{\infty}$ defined
by
\[T[n]=\left\{\begin{array}{ll}
1&\mathrm{if}\ s_n\in A \\
-1&\mathrm{if}\ s_n\in B \\
0&\mathrm{if}\ s_n\not\in A\cup B\end{array}\right.\] We identify
each disjoint pair with the corresponding sequence.
\end{notation}

Resource-bounded genericity for disjoint pairs is the natural extension of the concept introduced for languages by Ambos-Spies, Fleischhack and Huwig
\cite{AmFlHu87}.

\begin{definition}
A {\sl condition\/} $C$ is a set $C\subseteq \{-1,0,1\}^*$. A
{$t(n)$-condition\/} is a condition $C\in\DTIME(t(n))$. A
condition $C$ is {\sl dense along a pair $(A, B)$\/} if there are
infinitely many $n\in\N$ such that $(A,B)[0 .. n-1]i \in C$ for
some $i\in\{-1,0,1\}$. A pair $(A, B)$ {\sl meets\/} a condition
$C$ if $(A,B)[0 .. n-1] \in C$ for some $n$. A pair $(A, B)$ is
{\sl $t(n)$-generic\/} if $(A, B)$ meets every $t(n)$-condition
that is dense along $(A, B)$.
\end{definition}

We first prove that generic pairs are inseparable.

\begin{theorem}\label{the51}
Every $t(\log n)$-generic disjoint pair is $\TIME(t(n))$-inseparable.
\end{theorem}

We can now relate genericity in disjNP and inseparable pairs as
follows.

\begin{corollary}\label{cor52}
If disjNP contains a $2^{(\log n)^k}$-generic pair for every $k\in \N$, then disjNP contains a $\TIMEdnk$-inseparable pair for every $k\in \N$.
\end{corollary}

Resource-bounded measure on classes of disjoint pairs is the
natural extension of the concept introduced for languages by Lutz
\cite{AEHNC}, and is defined by using martingales on a
three-symbol alphabet as follows.

\begin{definition}\begin{enumerate}\item A {\sl pair martingale\/} is a function $d: \{-1,0,1\}^*\to [0, \infty)$ such that for every $w\in \{-1,0,1\}^*$
\[ d(w)=\frac{1}{4}d(w0)+\frac{3}{8}d(w1)+\frac{3}{8}d(w(-1)).\]
\item A pair martingale $d$ {\sl succeeds on a pair $(A, B)$\/} if
$\limsup_{w\to (A,B)}d(w)=\infty$. \item A pair martingale $d$
{\sl succeeds on a class of pairs $X\subseteq
\{-1,0,1\}^{\infty}$\/} if it succeeds on each $(A,B)\in X$.
\end{enumerate}
\end{definition}

Our intuitive rationale for the coefficients in part 1 of this
definition is the following. We toss one fair coin to decide
whether $s_{|w|}\in A$ and another to decide whether $s_{|w|}\in
B$. If both coins come up heads, we toss a third coin to break the
tie. The reader may feel that some other coefficients, such as
$\frac{1}{3}, \frac{1}{3}, \frac{1}{3}$ are more natural here.
Fortunately, a routine extension of the main theorem of
\cite{Breutzmann1999} shows that the value of $\mu(\disjNP\mid
\disjEXP)$ will be the same for {\sl any\/} choice of three
positive coefficients summing to 1.

 When restricting martingales to those computable within a
certain resource bound, we obtain a resource-bounded measure that
is useful within a complexity class. Here we are interested in the
class of disjoint EXP pairs, disjEXP.

\begin{definition}
\begin{enumerate}
\item Let $\ptwo$ be the class of functions that can be computed
in time $2^{(\log n)^{O(1)}}$.
 \item A class of pairs $X\subseteq \{-1,0,1\}^{\infty}$ has
{\sl $\ptwo$-measure 0\/} if there is a martingale $d\in\ptwo$
that succeeds on $X$. \item $X\subseteq \{-1,0,1\}^{\infty}$ has
{\sl $\ptwo$-measure 1\/} if $X^c$ has $\ptwo$-measure 0. \item A
class of pairs $X\subseteq \{-1,0,1\}^{\infty}$ has {\sl measure 0
in disjEXP\/}, denoted $\mu(X\mid \disjEXP)=0$, if $X\cap
\disjEXP$ has $\ptwo$-measure 0. \item $X\subseteq
\{-1,0,1\}^{\infty}$ has {\sl measure 1 in disjEXP\/} if $X^c$ has
measure 0 in disjEXP.
\end{enumerate}
\end{definition}

It is easy to verify that $\ptwo$-measure is nontrivial on disjEXP
(as proven for languages in \cite{AEHNC}).

In the following we consider the hypothesis that disjNP does not
have measure 0 in disjEXP (written   $\mu(\disjNP\mid \disjEXP)\ne
0$). We start by proving that this hypothesis is at least as
strong as the well studied  $\mu(\NP\mid \EXP)\ne 0$ hypothesis.

\begin{theorem}\label{the53} If $\mu(\disjNP\mid \disjEXP)\ne 0$ then $\mu(\NP\mid \EXP)\ne 0$.
\end{theorem}

We finish by relating measure and genericity for disjoint pairs.

\begin{theorem}\label{the54}
If $\mu(\disjNP\mid \disjEXP)\ne 0$, then disjNP contains a
$2^{(\log n)^k}$-generic pair for every $k\in \N$.
\end{theorem}

\begin{figure}

\includegraphics[height=12cm]{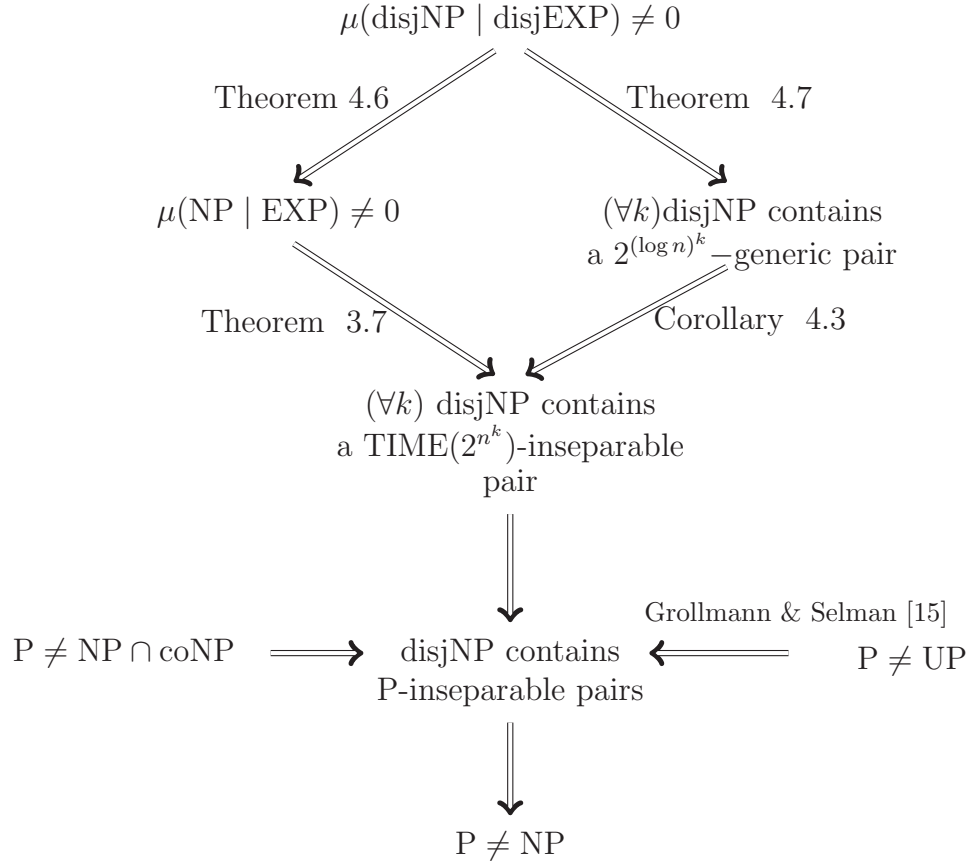}

\caption{Relations among some strong hypotheses.}
\end{figure}

\section{Oracle Results}\label{sec6}
\newcommand{\tuple}[1]{\langle #1 \rangle}

All the techniques in this and related papers relativize,
that is they hold when all machines involved have access to
the same oracle $A$. In this section we give relativized
worlds where the converses of most of the results in this
paper, as expressed in Figure \arabic{fig2}, do not hold.
Since the implications trivially all hold in any
relativized world where $\P = \NP$~\cite{BaGiSo75}, one
cannot use relativizable techniques to settle these
converses.

We'll work our way from the bottom up of Figure
\arabic{fig2}.
\begin{theorem}[Homer-Selman \cite{HomSel92}, Fortnow-Rogers~\cite{ForRog02}]
There exists oracles $A$ and $B$ such that
\begin{itemize}
\item $\P^A\neq \NP^A$ and $\disjNP^A$ does not contain $\P^A$-inseparable pairs.
\item $\P^B=\NP^B\cap\coNP^B=\UP^B$ and $\disjNP^B$ does contain
$\P^B$-inseparable pairs.
\end{itemize}
\end{theorem}

\begin{theorem}\label{the63} There exists an oracle $C$ such that
$\P^C\neq\UP^C$ but $\NP^C$ is contained in $\TIME^C(n^{O(\log
n)})$. In particular this means that relative to $C$, $\disjNP$
contains $\P$-inseparable pairs but there is a $k$ (and in fact
any real $k>0$) such that $\disjNP$ has no
$\TIME(2^{n^k})$-inseparable pairs.
\end{theorem}

\begin{theorem}\label{the64} There exists a relativized world $D$, relative to which for all
$k$, $\disjNP$ contains a $\TIME(2^{n^k})$-inseparable pair but
$\mu(\NP|\EXP)=0$ and $\disjNP$ does not contain a $2^{(\log
n)^k}$-generic pair.
\end{theorem}

\begin{theorem}\label{the65}
There exists an oracle $E$ relative to which for all $k$,
$\disjNP$ contains a $2^{(\log n)^k}$-generic pair but
$\mu(\disjNP|\disjEXP)=0$.
\end{theorem}

\begin{conjecture}
There exists an oracle $H$ relative to which $\mu(\NP|\EXP)\neq 0$ but \newline $\mu(\disjNP|\disjEXP)=0$.
\end{conjecture}

Let $K$ be a $\PSPACE$-compete set, $R$ be a ``random'' oracle and
let
\[H=K\oplus R=\{\tuple{0,x}\ |\ x\in K\}\cup\{\tuple{1,y}\ |\ y\in
R\}.\]

Kautz and Miltersen show in \cite{KauMil94} that relative to $H$,
$\mu(\NP|\EXP)\neq 0$. Kahn, Saks and Smyth \cite{KaSaSm00}
combined with unpublished work of Impagliazzo and Rudich show that
relative to $H$ there is a polynomial-time algorithm that solves
languages in $\NP\cap\coNP$ on average for infinitely-many lengths
which would imply $\mu(\NP\cap\coNP|\EXP)=0$ relative to $H$. We
conjecture that one can modify this proof to show
$\mu(\disjNP^H|\disjEXP^H)=0$.


\bibliographystyle{plain}
\bibliography{../../biblios/todo}

\end{document}